\documentclass[prl,floatfix,twocolumn,superscriptaddress,showpacs,amsmath,amssymb,showpacs]{revtex4-1}

\usepackage{graphicx}
\usepackage{epstopdf}
\usepackage{epsfig}
\usepackage{epsf}
\usepackage{url}
\usepackage[USenglish]{babel}
\usepackage{hyperref}
\def\bcen{\begin{center}}
\def\ecen{\end{center}}
\allowdisplaybreaks
\renewcommand\[{\begin{equation}}
\renewcommand\]{\end{equation}}
\usepackage{verbatim}
\usepackage{natbib}
\usepackage{amsmath, nccmath}
\usepackage[export]{adjustbox}
\usepackage{dcolumn}
\usepackage{bm}
\usepackage{bbm}
\usepackage{lipsum}
\usepackage{tikz-feynman}

\begin{document}
\title{Quaternary borocarbides: a testbed for DFT for superconductors}
\author{Viktor Christiansson}
\affiliation{Department of Physics, University of Fribourg, 1700 Fribourg, Switzerland}
\author{Philipp Werner}
\affiliation{Department of Physics, University of Fribourg, 1700 Fribourg, Switzerland}

\begin{abstract}

Using {\it ab-initio} density functional theory for superconductors (SCDFT), we systematically study the quaternary borocarbides $RM_2$B$_2$C. Treating the retarded (frequency-dependent) interaction $W(\omega)$ within the random-phase approximation (RPA), we find good agreement with experiments for the calculated superconducting critical temperature $T_c$ in the nonmagnetic Ni- and Pd-based compounds.
We argue that the problem of accurately placing the $f$-bands within DFT, and possibly the lack of an explicit magnetic pair-breaking mechanism, explain the difficulties of SCDFT in reproducing $T_c$ in members of the magnetic $R$Ni$_2$B$_2$C series ($R$ rare-earth with partially filled $4f$).
While the calculated $T_c$ is overestimated, SCDFT qualitatively captures the experimentally observed trend along the rare earth series, which indicates that the electron-phonon couplings and dynamically screened interactions have a significant effect on $T_c$.

\end{abstract}

\maketitle

{\it Introduction}
The discovery of superconductivity in a quaternary borocarbide Y-Ni-B-C compound \cite{Nagarajan1994,Cava1994} 
quickly lead to the synthesis and characterization of a large number of superconducting and non-superconducting members of this family \cite{Tominez2000,Niewa2011}.
The superconducting compounds with general formula $RM_2$B$_2$C are limited to the $M=\text{Ni}$ \cite{Cava1994,Lai1995,Lynn1997}, Pd \cite{Cava1994c,Jiang1995,Sarrao1994}, and Pt \cite{Cava1994b,Buchgeister1995,Sarrao1994} group and Ru \cite{Hsu1998}, with $R=\text{Sc}$, Y, Th, or a rare-earth (depending on $M$).
The highest critical temperature $T_c$ in the family was reported for YPd$_2$B$_2$C ($\sim23$~K) \cite{Cava1994c} and,
as a general observation, $T_c$ displays systematic changes depending on the in-plane $M$-$M$ distance (or $R$ ionic radius) \cite{Lai1995,Muller2001}.
A further intriguing and much studied aspect is the existence of Ni-based $R$Ni$_2$B$_2$C with comparable superconducting (increasing) and magnetic (decreasing) ordering temperatures when moving along the rare-earth series from Tb to Lu \cite{Cava1994,Muller2001}.
Structurally, the borocarbides are filled ThCr$_2$Si$_2$-type layered systems \cite{Siegrist1994} with space group $I4/mmm$ (139).

The large number of characterized isostructural compounds provides an ideal testing ground for theoretical predictions of superconductivity, since it enables the systematic alteration of physical properties. Often, the critical temperature $T_c$ of conventional superconductors is estimated using the phenomenological McMillan formula \cite{McMillan1968,Allen1975}. Here, the electron-phonon coupling constant $\lambda$, calculated within density functional theory (DFT) \cite{Hohenberg1964,Kohn1965}, is combined with a typically {\it ad hoc} value for the pseudopotential $\mu^*$ (representing the effective repulsive electron-electron interaction) to reproduce the experimental $T_c$. Indeed, most theoretical studies estimating the critical temperature of borocarbides have used this type of methodology  \cite{Reichardt2005,Weber2014,Tutuncu2015,Uzunok2015}. A notable exception is the work by Kawamura and co-workers \cite{Kawamura2017}, which employed the parameter-free density functional theory for superconductors (SCDFT) \cite{Oliveira1988,Luders2005} and found close agreement with the experimentally reported $T_c$ for YNi$_2$B$_2$C. They pointed out that reproducing this $T_c$ with the McMillan equation would require an unusually small $\mu^*$.

Here, we use the quaternary borocarbide compounds to systematically test the predictive power of SCDFT. 
We find a remarkably good agreement, both qualitatively and quantitatively, for the series of non-magnetic Ni- and Pd-based compounds, while the Pt- and magnetic Ni-based systems display larger deviations.
Our results furthermore demonstrate the usual difficulty in including localized $4f$ states in a DFT description and point to the need for incorporating an explicit magnetic pair-breaking mechanism. However, our analysis also indicates that pair-breaking alone cannot explain the $T_c$ changes over the rare-earths, and that dynamical screening plays a role.

\begin{figure}[t]
\begin{centering}
\includegraphics[width=0.98\columnwidth]{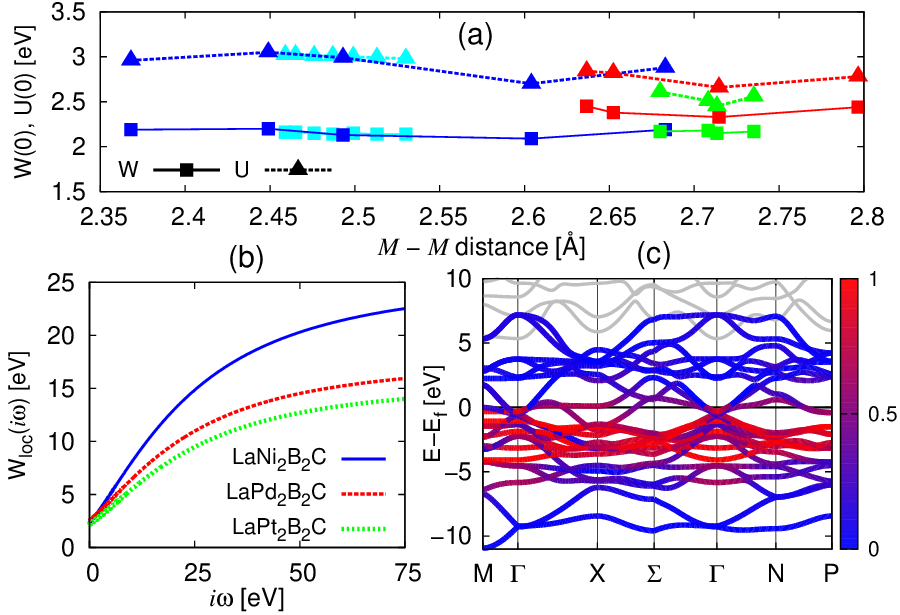}
\end{centering}
\caption{
(a) Local (${\bf R}=0$) static screened $W(\omega=0)$ and effective bare $U_\textrm{cRPA}(\omega=0)$ interaction for the $M$--$d$ states with $M=\textrm{Ni}$ (blue; magnetic turquoise), Pd (red), and Pt (green).
(b) Representative frequency-dependence of $W_\text{loc}(i\omega)$, shown for the $M$--$d$ orbitals of the La-based compounds. (c) DFT band structure for ScNi$_2$B$_2$C. The colored bands are the disentangled (Wannier) band structure with the projected Ni-$3d$ orbital character indicated by the colorbar.
\label{Fig:interaction}
}
\end{figure}

{\it Method.}
We start from DFT calculations in the normal state for each material, relaxing the ionic positions while fixing the lattice constants to the experimental values. Using this relaxed structure we calculate the electron-phonon matrix elements $g_{\lambda_{\bf q}}^{n{\bf k},n'{\bf k'}}$ using density functional perturbation theory \cite{Baroni2001}, as implemented in \textsc{Quantum} ESPRESSO \cite{Giannozzi2009,Giannozzi2017}, with a $6\times6\times6$ ${\bf q}$-grid. 
To compute the electron-electron interaction, we define a low-energy model using maximally localized Wannier functions \cite{Marzari1997,Mostofi2008} (with initial $R$--$d$, $M$--$d$, B--$p$ and C--$p$ projections), and calculate the corresponding frequency dependent screened interaction $W_{\bf k}(\omega)$ within RPA \cite{Pines1963} using the SPEX code \cite{Friedrich2010}. We also calculate the effective bare interaction using constrained RPA (cRPA) \cite{Aryasetiawan2004}, where we exclude screening only from within the $M$--$d$ manifold, which gives the main contribution to the density of states (DOS) close to the Fermi energy [Fig.~\ref{Fig:interaction}(c)].

In SCDFT the superconducting critical temperature $T_c$ is calculated from the vanishing of the gap function $\Delta$ in the self-consistent solution of the gap equation
\begin{equation}\label{eq:gapequation}
\Delta_{n\mathbf{k}}=-\mathcal{Z}_{n\mathbf{k}}\Delta_{n\mathbf{k}}-\frac{1}{2}\sum_{n'\bf{k}'}\mathcal{K}_{n{\mathbf{k}},n'\mathbf{k'}}\frac{\tanh[(\beta/2)E_{n'\bf{k}'}]}{E_{n'\mathbf{k}'}}\Delta_{n'\mathbf{k}'}.
\end{equation}
Here $E_{n{\bf{k}}}=\sqrt{\varepsilon_{n{\bf{k}}}^2+\Delta_{n{\bf{k}}}^2}$ with $\varepsilon_{n{\bf{k}}}$ the Kohn-Sham eigenenergies, $\beta$ is the inverse temperature, and we only consider states within the low-energy window. 
$\mathcal{Z}$ and $\mathcal{K}$ are the SCDFT exchange correlation kernels describing the effects of electron-phonon and electron-electron interactions.
We solve the energy-averaged form \cite{Marques2005} of Eq.~\eqref{eq:gapequation} and take the phononic kernels $\mathcal{Z}^\textrm{ph}$ and $\mathcal{K}^\textrm{ph}$ developed by L\"uders, Marques, and co-workers \cite{Luders2005,Marques2005}, defined in terms of the Eliashberg function $\alpha^2F(\omega)$ calculated from the couplings $g_{\lambda_{\bf q}}^{n{\bf k},n'{\bf k'}}$ and phonon frequencies $\omega_{\lambda{\bf q}}$ \cite{Allen1972}.
Finally, to treat the (retarded) electron-electron interaction, we consider the frequency-dependent contribution from $W_{\bf k}(\omega)$ to the electronic kernel $\mathcal{K}^\textrm{el}$ proposed by Akashi and Arita \cite{Akashi2013,Akashi2014}.
In the present formalism we therefore treat both the usual Coulomb repulsion and electron-phonon mediated attraction, while also allowing for the possibility of an enhanced retardation effect \cite{Morel1962,Akashi2014}.
The details of our implementation can be found in Ref.~\onlinecite{Christiansson2022} and a summary of the formalism in the Supplemental Material (SM). Since the DOS changes rapidly around the Fermi energy for several considered systems, we do not assume $N(\varepsilon)$ to remain constant in the calculation of $\mathcal{Z}^\textrm{ph}$. Instead, we adopt the approach in Ref.~\cite{Akashi2013b} for the averaging over isoenergetic surfaces
\footnote{We have also tested the asymmetrical form of the electron-phonon $\mathcal{Z}^\textrm{ph}$ kernel proposed in Ref.~\cite{Akashi2013b} which gives only minor differences of $\sim1$-2 \%.}. 
This treatment leads to $T_c$ changes of 1-2~K in some compounds, compared to an assumed constant DOS.

\begin{figure}[t]
\begin{centering}
\includegraphics[width=0.98\columnwidth]{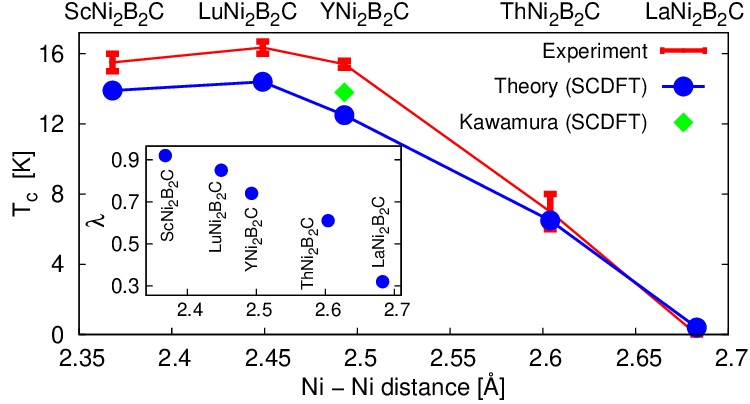}
\end{centering}
\caption{$T_c$ as a function of Ni-Ni distance for the non-magnetic $R$Ni$_2$B$_2$C borocarbides, calculated within SCDFT and compared to experimental values (see Ref.~\onlinecite{Niewa2011} and SM). The errorbars for the experimental data indicate the range of reported $T_c$. The green diamond is the SCDFT result from Ref.~\onlinecite{Kawamura2017}. The inset shows the calculated electron-phonon coupling constant $\lambda$.
\label{Fig:Ni_Tc}}
\end{figure}

{\it Non-magnetic $R$Ni$_2$B$_2$C.}
We first focus on the Ni-based members of the family. The calculated $T_c$ are shown in Fig.~\ref{Fig:Ni_Tc} as a function of Ni-Ni distance for the non-magnetic systems. We find a good agreement with experiment, reproducing the experimentally observed ridge (or slight peak) from Sc to Lu, followed by a rapid decrease with further increasing distance.
While experimentally, the La-based compound is nonsuperconducting, in our calculations we find a non-zero $T_c$ ($\lesssim 0.4$ K), which is however consistent within the expected accuracy. 
Our results also compare well with the previously reported SCDFT result (green diamond) for YNi$_2$B$_2$C by Kawamura {\it et al.} \cite{Kawamura2017}, although our $T_c$ is $\sim 1$~K lower 
\footnote{This difference is likely due to differences in the computational details, e. g., our energy-averaging which does not consider the gap function anisotropy, the use of relaxed versus experimental lattice constants, and the electron interaction calculated with an all-electron versus pseudopotential code.}.

The effective bare interactions for the Ni $3d$ electrons, calculated using cRPA, are shown in Fig~\ref{Fig:interaction}(a) (blue triangles) and listed in Table~\ref{Table:parameters}. The bare values are somewhat smaller than for other Ni compounds (e.g., elemental Ni \cite{Miyake2009} and various nickelate superconductors \cite{Petocchi2020,Christiansson2023a,Christiansson2023b}), and the Hund's coupling is screened only very weakly from the bare value $J\approx0.8$ eV to $\tilde{J}\approx0.7$ eV.
The bandwidth of the mostly Ni-derived bands, shown for a representative case in Fig.~\ref{Fig:interaction}(c), is on the order of $\sim 5$~eV, placing the borocarbides into the weak-to-moderate correlation regime. Hence, we expect that RPA can provide a reasonable description.

While our calculations take into account frequency- and momentum-dependent screening, it is instructive to consider the screening of the local interaction $W_\text{loc}$. Figure~\ref{Fig:interaction}(a) shows that the static $W_\textrm{loc}(\omega=0)$ (blue squares) remains approximately constant over the range of Ni-based borocarbides, and that the $d$ states screen the bare values only relatively weakly. 
Since $\lambda$ monotonically decreases, but there is no simultaneous reduction in the static local repulsion, it is clear that we need to consider the ${\bf k}$- and $\omega$-dependence to explain the comparable $T_c$ values for Sc, Lu and Y.
Explicitly treating the retardation effect, encoded in the frequency dependent electron interaction, is particularly important, since the use of the static interaction leads to an underestimation of $T_c$ by a factor of at least 3. The extreme case is YNi$_2$B$_2$C, with $T_c^\textrm{exp}/T_c^\textrm{stat}\approx 8$ (see SM). Nonlocal dynamical screening is therefore primarily responsible for the material-dependent enhancement of $T_c$.

\begin{table}
\caption{Calculated electron-phonon coupling constant $\lambda$ and $\omega_\textrm{ln}$ in K, used in the McMillan equation, and the effective bare interaction $U_\textrm{cRPA}(\omega=0)$ in eV for the $M$--$d$ states, calculated within cRPA.
\label{Table:parameters}
}
\setlength{\tabcolsep}{4.2pt} 
\renewcommand{\arraystretch}{1.21} 
\begin{tabular}{|c|c c c||c|c c c|}
\hline

\hline

\multicolumn{1}{c}{}  & \multicolumn{1}{c}{$\lambda$} & \multicolumn{1}{c}{$\omega_\textrm{ln}$} & \multicolumn{1}{c}{$U$} & \multicolumn{1}{c}{}  & \multicolumn{1}{c}{$\lambda$} & \multicolumn{1}{c}{$\omega_\textrm{ln}$} & \multicolumn{1}{c}{$U$} \\ 
\hline

\hline
\hline
LaNi$_2$B$_2$C & 0.32 &  271  &  2.9 & 
TmNi$_2$B$_2$C & 0.81 &  276  &  3.0 \\

ThNi$_2$B$_2$C & 0.61 &  266  &  2.7 &
ErNi$_2$B$_2$C & 0.79 &  279  &  3.0 \\

YNi$_2$B$_2$C & 0.74 &  299  &  3.0 & 
HoNi$_2$B$_2$C & 0.76 &  281  &  3.0 \\

LuNi$_2$B$_2$C & 0.85 &  267  & 3.1 & 
DyNi$_2$B$_2$C & 0.68 &  283  &  3.0 \\

ScNi$_2$B$_2$C & 0.92 &  271  &  3.0 & 
TbNi$_2$B$_2$C & 0.61 & 295  &  3.0 \\

YbNi$_2$B$_2$C & 0.84 &  272  &  3.0 &
GdNi$_2$B$_2$C & 0.58 &  295  & 3.0 \\

\hline
\hline

LuPd$_2$B$_2$C & 1.93 &  155  & 2.8 &
YPt$_2$B$_2$C & 1.13 &  207  & 2.6 \\

YPd$_2$B$_2$C & 1.35 &  222  &  2.8 &
ThPt$_2$B$_2$C & 0.92 &  206  &  2.5 \\

ThPd$_2$B$_2$C & 0.90 &  236  & 2.7 &
PrPt$_2$B$_2$C & 0.85 & 242 & 2.5 \\

LaPd$_2$B$_2$C & 0.43 & 247 & 2.8  &
LaPt$_2$B$_2$C & 0.72 & 255  &  2.6 \\

\hline

\end{tabular}
\end{table}

We should point out that while there are no adjustable parameters in the method, our accuracy is limited by the numerical convergence; the electron-phonon coupling can depend substantially on how well-converged the phonon calculations are with respect to the ${\bf k}$- and ${\bf q}$-grids \cite{Savrasov1996,Koretsune2017}.
We should keep this in mind even before considering the inherent approximations of the formalism (DFT band structure, energy-averaging, bubble approximation in RPA, ...) which are harder to estimate and can result in appreciable differences, as discussed in Ref.~\onlinecite{Christiansson2022}. 
Nevertheless, our results show the importance of using {\it ab initio} calculated interaction strengths for predictions of $T_c$.
In fact, to reproduce our calculated critical temperatures using the phenomenological McMillan equation and $\lambda$ and $\omega_\text{ln}$ from Table~\ref{Table:parameters}, one would have to choose $\mu^*$ in a range from $\sim0.08$ (La) to $\sim0.13$ (Sc).

\begin{figure}[t]
\begin{centering}
\includegraphics[width=0.97\columnwidth]{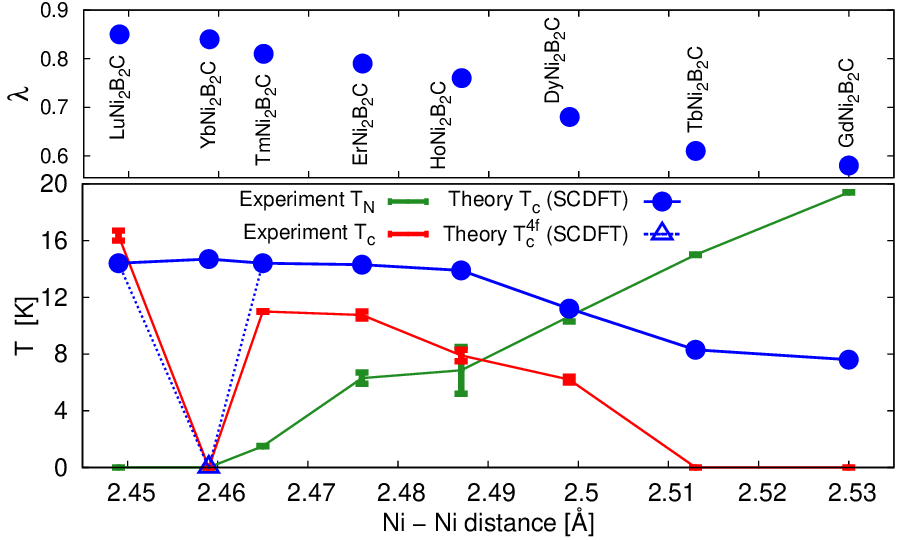}
\end{centering}
\caption{
Same as Fig.~\ref{Fig:Ni_Tc} for the magnetic Ni-based $R$Ni$_2$B$_2$C ($R=\textrm{rare-earth}$) with the partially filled $4f$ states placed in the core (Lu is included again for comparison). Also shown is the magnetic ordering temperature $T_N$ (green line) and the calculation with $4f$ states for YbNi$_2$B$_2$C (blue triangle).
\label{Fig:magnetic_Ni_Tc}}
\end{figure}

{\it Magnetic $R$Ni$_2$B$_2$C.}
Next, we consider the magnetic Ni-based members shown in Fig.~\ref{Fig:magnetic_Ni_Tc} ($R$ is in the latter half of the rare-earth series). Experimentally, with the exception of Yb, the measured $T_c$ systematically decreases as we go down the series from nonmagnetic Lu. At the same time there is a systematic increase in the magnetic ordering temperature $T_N$, leading to a dome-shaped coexistence region of the two orders.
The opposite trends are naturally explained by the pair-breaking caused by the magnetic $4f$ states \cite{Eisaki1994,Muller2001}, which becomes more significant towards the middle of the rare-earth series.
It is, however, less clear how much of the total suppression of $T_c$ can be attributed to the pair-breaking and what part could already be expected from a reduction of the phonon-mediated attraction, or increase in the Coulomb repulsion (nonmagnetic La is also nonsuperconducting).
While the experimental $T_c$ decreases with an increasing de Gennes (dG) factor ($\sim$ exchange interaction \cite{Muller2001}) and the systematic changes have therefore been attributed to magnetic pair-breaking \cite{Eisaki1994}, one finds a different scaling depending on the ratio between $T_c$ and $T_N$  \cite{Cho1996}.
It has been argued that this cannot be explained by a simple change in the DOS at the Fermi energy and that other nonmagnetic effects should play a role \cite{Muller2001}.
Also, an extrapolation of the dG scaling to TbNi$_2$B$_2$C predicts a $T_c$ of 4 K \cite{Tominez2000}, while this material is not superconducting.

Complicating things on the theory side, the series of rare-earths has partially filled $4f$ subshells. This causes issues within DFT, as the $f$ bands are in general not sufficiently separated and cross the Fermi energy. 
As a consequence, their effect on the low-energy physics is overestimated, and fully suppresses the superconducting state (we will show this explicitly for YbNi$_2$B$_2$C).
We therefore focus first on calculations where we neglect the hybridization between $4f$ and conduction electrons by treating the $4f$ states as part of the core. By ignoring these states, and an explicit magnetic pair-breaking mechanism, we can investigate other possible mechanisms playing a role in the experimentally observed $T_c$ trend.

Based on the electron-phonon coupling constants $\lambda$ (Fig.~\ref{Fig:magnetic_Ni_Tc} and Table~\ref{Table:parameters}) we would naively expect a range of $T_c$ comparable to the Lu and Th compounds above, which is confirmed by the full SCDFT results in Fig.~\ref{Fig:magnetic_Ni_Tc}.
We find an almost constant $T_c$ up to $R=\textrm{Er}$, despite the (modest) changes in $\lambda$, in reasonable agreement with experiment (Lu has a filled $4f$ subshell). This is in contrast to the $T_c$ calculated using a static-only electronic kernel, where the severely underestimated values monotonically decrease over the full series, except for Tm and Er (see SM), where $T_c$ again remains constant. These results indicate that the plateau found both experimentally and theoretically can be qualitatively explained by the electron-phonon coupling and screened interaction (although the $4f$ states would be required for a quantitative agreement).
The jump in the experimental $T_c$ at Ho is however not captured, which is not too surprising, since the magnetic and superconducting ordering temperatures are now comparable ($T_c/T_N\sim1$) and magnetic ordering (and pair-breaking) should become relevant.  
Moving on to Dy, SCDFT predicts a decrease of $T_c$ in agreement with that seen experimentally, albeit the absolute value of $T_c$ is overestimated. This change is expected, considering the decrease in $\lambda$ and the relatively constant screened interaction (Figs.~\ref{Fig:magnetic_Ni_Tc} and \ref{Fig:interaction}). Also for the end-members Tb and Gd we obtain a further decrease, although at this point calculations neglecting the $f$ states are clearly insufficient, and cannot explain the experimentally observed complete suppression of superconductivity.

While the evidence is not definitive, due to a number of approximations, our theoretical results suggest that several of the late rare-earth borocarbides (with $T_c>T_N$) can be qualitatively understood within a phonon mechanism and by considering the (dynamically) screened Coulomb interaction. These effects appear to be important also for understanding the earlier rare-earths where, although magnetic pair-breaking dominates, the dG scaling does not fully account for TbNi$_2$B$_2$C.
The remaining discrepancy to experiment, however, indicates that the magnetic $4f$ states should be responsible for most of the decrease in $T_c$ throughout the series, consistent with the current understanding \cite{Muller2001}.
In the future, it would be interesting to explore if a more accurate treatment of the $4f$ states would be sufficient to fully reproduce the experimentally observed suppression of $T_c$  within SCDFT. Alternatively, an explicit magnetic pair-breaking mechanism might have to be introduced, e.~g. based on the spin-fluctuation kernel proposed in Ref.~\onlinecite{Essenberger2014}.

Finally, we return to the outlier YbNi$_2$B$_2$C. Experimentally it is found to be a heavy Fermion system \cite{Yatskar1996}, and the hybridization with the $4f$ states should play a major role in the suppression of superconductivity.
Including the $4f$ states in the electronic kernel does in fact result in a complete suppression ($T_c=0$) [triangle in Fig.~\ref{Fig:magnetic_Ni_Tc}]. While our DFT treatment of the $4f$ states is clearly insufficient, this is an encouraging indication that SCDFT could be capable of capturing also this aspect.

\begin{figure}[t]
\begin{centering}
\includegraphics[width=0.98\columnwidth]{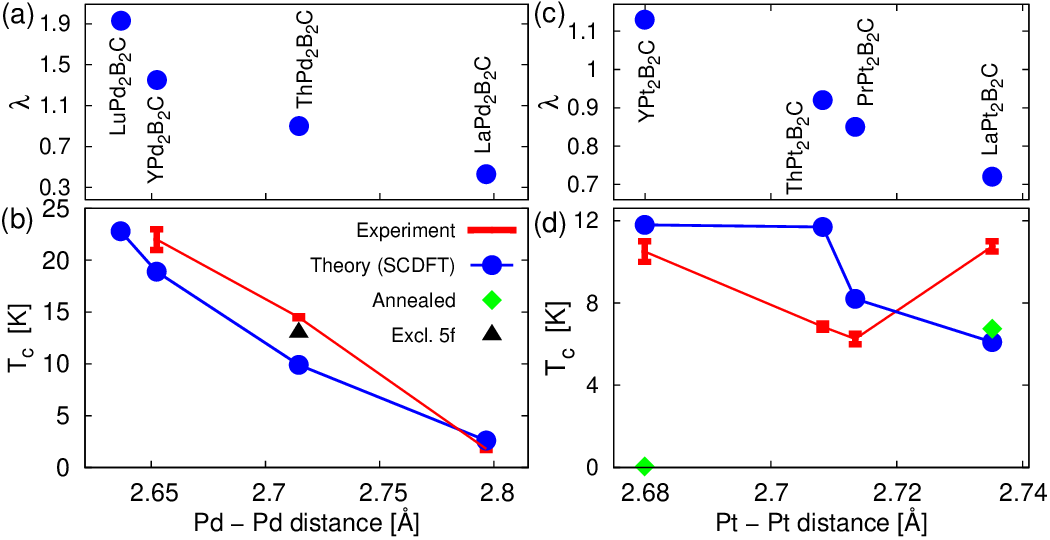}
\end{centering}
\caption{
Same as Fig.~\ref{Fig:Ni_Tc} for Pd- (a,b) and Pt-based (c,d) $RM_2$B$_2$C. In (b) we indicate ThPd$_2$B$_2$C with excluded $5f$ (black triangle) and in (d) we show the experimental $T_c$ after annealing in $R$Pt$_2$B$_2$C for $R=\textrm{Y}$ and La (green diamonds).
\label{Fig:Pd_Tc}}
\end{figure}

{\it Other borocarbides.}
Although experimentally, the Pd- and Pt-based compounds are generally metastable or multiphase \cite{Niewa2011}, we consider here only on the stoichiometric materials.
Starting with $M=\textrm{Pd}$ in Fig.~\ref{Fig:Pd_Tc}(a,b), the qualitative trend is reproduced well for the experimentally studied materials (Y, Th, and La), following the trend expected from $\lambda$. $R=\textrm{Th}$ shows the largest error compared to experiment, with DFT placing the unoccupied $5f$ states in the vicinity of the Fermi energy ($\sim 2$~eV above). 
Since they are well-separated in this case, we define a second low-energy space, where we explicitly exclude the unoccupied $f$ orbitals. 
While the DOS and disentangled band structure up to the $f$ states remain similar, the exclusion of the $f$ states from the electronic kernel has the effect of lifting $T_c$ by a few K, bringing our results into closer agreement with the experiments.

In contrast to the compounds discussed so far, the $T_c$ values computed for the Pt-based systems in Fig.~\ref{Fig:Pd_Tc}(c,d) are found to considerably deviate from the experimentally reported critical temperatures (PrPt$_2$B$_2$C hosts a partially filled $4f$ subshell, which is placed in the core). In particular, we do not capture the counterintuitive valley-like shape of the $T_c$.
On the other hand, annealing the samples leads to a reduction of the experimental $T_c$ to $\sim7$~K in LaPt$_2$B$_2$C  \cite{Cava1994b} (green diamond in Fig.~\ref{Fig:Pd_Tc}), in much better agreement with our theoretical prediction, whereas annealed YPt$_2$B$_2$C is nonsuperconducting \cite{Cava1994b}.
The experimental uncertainty makes it hard to draw firm conclusions on the accuracy of SCDFT in this case.

Regarding the multiphased YRu$_2$B$_2$C system \cite{Hsu1998}, let us note that our phonon calculations predict the stoichiometric compound to be unstable, although the total contribution of these modes to $\alpha^2F(\omega)$ is relatively small (see SM). Removing them, we calculate a $T_c$ of 17 K, compared to the experimental value of 10 K \cite{Hsu1998}.
The high theoretical $T_c$ results from the relatively large $\lambda=1.3$ (due to the softened phonon modes) and a small static repulsion $W_{\textrm{Ru-}d}(0)=1.2$ eV. 
Further experiments on this compound would be helpful.

The local static interaction strengths for Pd and Pt shown in Fig.~\ref{Fig:interaction} and Table~\ref{Table:parameters} are comparable to those in the Ni-based systems. As expected for more delocalized $4d$ and $5d$ states, however, the bare Coulomb interaction (infinite-frequency limit) is systematically lowered [Fig~\ref{Fig:interaction}(b)], which reduces the retardation effect encoded in $\mathcal{K}^\textrm{el}$. This can in part explain the lower calculated $T_c$ for systems which have a comparable, or even larger, $\lambda$ than the Ni-based borocarbides.

On a final note, the $R=\textrm{Sc}$ and Lu members have so far not been successfully synthesized for either $M=\textrm{Pd}$ or Pt \cite{Cava1994b,Niewa2011}.
We performed phonon calculations, relaxing also the lattice parameters, and out of the four, only LuPd$_2$B$_2$C is found to be dynamically stable (the others display significant imaginary phonon frequencies). 
The calculated $\lambda$ is the highest among the considered systems, and SCDFT accordingly predicts that it would have the highest $T_c$ ($\gtrsim 23$ K) among the borocarbides [Fig.~\ref{Fig:Pd_Tc}(b)].

{\it Conclusions}
The calculated superconducting $T_c$ of the quaternary borocarbide superconductors from parameter-free SCDFT show a consistently good agreement with experiment over a wide range of compounds. 
We also predict that the not yet synthesized LuPd$_2$B$_2$C would display the highest $T_c$ among the borocarbide superconductors.
The largest discrepancies are found for the multiphased Pt- and Ru-based materials.
We furthermore demonstrated the limitations of the current SCDFT method when not all mechanisms relevant for the superconductivity (e. g. magnetic pair-breaking) are considered. However, we do find indications that a complete description of the changes in $T_c$ for the magnetic rare-earth systems also needs to take into account the electron-phonon and dynamical screening induced changes, since the latter consistently explain the qualitative trends.
Future work is needed to clarify if a more accurate treatment of the $f$ orbitals within SCDFT is sufficient to reproduce the complete suppression of $T_c$ observed experimentally in the Ni-based Tb, Gd, and Yb compounds, or if a magnetic pair-breaking kernel is required.

{\it Acknowledgments} We acknowledge support from the Swiss National Science Foundation via NCCR Marvel. The calculations have been performed on the beo05 cluster at the University of Fribourg. 

\bibliography{borocarbides}

\end{document}


\title{Quaternary borocarbides: a testbed for DFT for superconductors Supplemental Material}

\author{Viktor Christiansson}
\affiliation{Department of Physics, University of Fribourg, 1700 Fribourg, Switzerland}
\author{Philipp Werner}
\affiliation{Department of Physics, University of Fribourg, 1700 Fribourg, Switzerland}

\maketitle

\section{The SCDFT formalism}

Density functional theory for superconductors (SCDFT) \cite{Oliveira1988,Luders2005,Marques2005} is a parameter-free method which allows to calculate the superconducting critical temperature $T_c$ of realistic materials from first principles. 
Through the self-consistent solution of the gap equation 
\begin{equation}\label{eq:gap}
\Delta_{n\bf{k}}=-\mathcal{Z}_{n\bf{k}}\Delta_{n\bf{k}}-\frac{1}{2}\sum_{n'\bf{k}'}\mathcal{K}_{n{\bf{k}},n'\bf{k'}}\frac{\tanh[(\beta/2)E_{n'\bf{k}'}]}{E_{n'\bf{k}'}}\Delta_{n'\bf{k}'},
\end{equation}
with $E_{n{\bf{k}}}=\sqrt{\varepsilon_{n{\bf{k}}}^2+\Delta_{n{\bf{k}}}^2}$ and $\beta$ the inverse temperature, the critical temperature is estimated from the vanishing of the gap function $\Delta_{n{\bf{k}}}$.
Through the choice of the approximation for the exchange-correlation kernels $\mathcal{Z}$ and $\mathcal{K}$, one can include different mechanisms relevant for the description of the superconducting state, such as conventional electron-phonon pairing, (repulsive) electron-electron interactions and retardation effects \cite{Luders2005,Marques2005}, plasmon mediated attractions \cite{Akashi2013,Akashi2014}, or pairing mediated by spin-fluctuations \cite{Essenberger2014}. The details of our implementation have been presented in Ref.~\cite{Christiansson2022}, and in this section we briefly summarize the formalism.

We will consider the most commonly used electron-phonon kernels $\mathcal{Z}^\textrm{ph}$ and $\mathcal{K}^\textrm{ph}$ proposed by L\"uders, Marques and co-workers in Refs.~\cite{Luders2005,Marques2005} and the frequency dependent (dynamical) electronic kernel $\mathcal{K}^\textrm{el}$ proposed by Akashi and Arita  \cite{Akashi2013,Akashi2014}.
The latter proves to be important for the quantitative estimate of $T_c$ in the borocarbides. As discussed in the main text and further shown below, the use of only a static RPA interaction overestimates the electronic repulsion and leads to severely underestimated $T_c$ values. In our calculations, the exchange-correlation kernels entering Eq.~\eqref{eq:gap} are then $\mathcal{Z}=\mathcal{Z}^\textrm{ph}$ and $\mathcal{K}=\mathcal{K}^\textrm{ph}+\mathcal{K}^\textrm{el static}+\mathcal{K}^\textrm{el dyn}$, where the electronic kernel has been further split up into two parts depending on the static and dynamical interactions, respectively.

A direct solution of Eq.~\eqref{eq:gap} requires a very fine sampling of ${\bf k}$-points around the Fermi level \cite{Marques2005}, since especially the phononic part has a very strong energy dependence in a narrow energy region. We instead use the energy-averaged version of the formalism \cite{Marques2005}, 
\begin{equation}
\Delta(\varepsilon)=-\mathcal{Z}(\varepsilon)\Delta(\varepsilon)- \frac{1}{2}\int \mathrm{d} \varepsilon' N(\varepsilon')\mathcal{K}(\varepsilon,\varepsilon')\frac{\tanh[(\beta/2) E']}{E'} \Delta(\varepsilon'),
\label{eq:gap_enav}
\end{equation}
where the gap equation has been averaged over isoenergetic surfaces. The electron-phonon contribution to $\mathcal{K}$ is expressed as
\begin{align}
\mathcal{K}^\text{ph}(\varepsilon,\varepsilon')=\frac{2}{\tanh(\varepsilon\beta/2)\tanh(\varepsilon'\beta/2)}\frac{1}{N(0)} 
 \int  \mathrm{d}\omega \alpha^2F(\omega) \left[ I(\varepsilon,\varepsilon',\omega)-I(\varepsilon,-\varepsilon',\omega)\right],
\end{align}
with $N(0)$ the density of states at the Fermi energy, and where the function
\begin{equation}
I(\varepsilon,\varepsilon',\omega)= n_F(\varepsilon)n_F(\varepsilon')n_B(\omega) \left( \frac{e^{\beta \varepsilon}-e^{\beta(\varepsilon'+\omega)}}{\varepsilon-\varepsilon'-\omega} - \frac{e^{\beta \varepsilon'}-e^{\beta(\varepsilon+\omega)}}{\varepsilon-\varepsilon'+\omega} \right)
\end{equation}
is defined in terms of the Fermi-Dirac ($n_F(\varepsilon)$) and Bose-Einstein ($n_B(\omega)$) distributions.
The Eliashberg function
\begin{align}\label{Eq:a2F}
\alpha^2F(\omega) =& \frac{1}{N(0)}\sum_{\lambda,{\bf q}}\sum_{nn',{\bf k}}\left| g_{\lambda,{\bf q}}^{n{\bf k},n'{\bf (k+q)}} \right|^2 \delta(\varepsilon_{n{\bf{k}}})\delta(\varepsilon_{n' {\bf k}+{\bf q}})\delta(\omega-\omega_{\lambda{\bf q}})
\end{align}
is calculated from the phonon frequencies $\omega_{\lambda {\bf q}}$ for wave vector ${\bf q}$ and mode $\lambda$
and the matrix elements of the electron-phonon coupling constants
\begin{equation}
g_{\lambda,{\bf q}}^{n{\bf k},n'{\bf (k+q)}} = \frac{1}{\sqrt{2M\omega_{\lambda{\bf q}}}} \left \langle n',({\bf k}+{\bf q}) \right| \delta_{{\bf q}}^\lambda \, V^\text{KS} \left| n,{\bf k} \right\rangle.
\end{equation}
Here, $\delta_{{\bf q}}^\lambda \, V^\text{KS}$ denotes the variation of the Kohn-Sham potential with respect to the displacements and $M$ the atomic mass.

The second phononic contribution to the kernels, $\mathcal{Z}^\text{ph}$, can be expressed as
\begin{equation}\label{eq:Zph}
\mathcal{Z}^\text{ph}(\varepsilon)=-\frac{1}{\tanh(\beta/2\varepsilon)}\int \mathrm{d} \varepsilon' \frac{N(\varepsilon')}{N(0)} \int  \mathrm{d} \omega \alpha^2F(\omega) \left[ J(\varepsilon,\varepsilon',\omega)+J(\varepsilon,-\varepsilon',\omega)\right],
\end{equation}
with
\begin{equation}
J(\varepsilon,\varepsilon',\omega)= \tilde{J}(\varepsilon,\varepsilon',\omega)-\tilde{J}(\varepsilon,\varepsilon',-\omega),
\end{equation}
\begin{equation}
\tilde{J}(\varepsilon,\varepsilon',\omega)=-\frac{n_F(\varepsilon)+n_B(\omega)}{\varepsilon-\varepsilon'-\omega}
\left( \frac{n_F(\varepsilon')-n_F(\varepsilon-\omega)}{\varepsilon-\varepsilon'-\omega} - \beta n_F(\varepsilon-\omega)n_F(-\varepsilon+\omega) \right).
\end{equation}
Here, unlike in our previous calculations in Ref.~\cite{Christiansson2022}, we do not assume $N(\varepsilon)$ to be constant in the narrow energy region around the Fermi level relevant for the phonons. Following Ref.~\cite{Akashi2013b}, this leads to the extra $\frac{N(\varepsilon')}{N(0)}$ factor in Eq.~\eqref{eq:Zph}.

Finally, the static contribution to the electronic interaction kernel,
\begin{equation}
\mathcal{K}^\textrm{el static}(\varepsilon,\varepsilon')=\frac{1}{N(\varepsilon)N(\varepsilon')}\sum_{n{\bf{k}},n'{\bf{k}'}}\delta(\varepsilon-\varepsilon_{n{\bf k}})\delta(\varepsilon'-\varepsilon_{n {\bf k'}})W_{n{\bf{k}},n'{\bf{k}'}}(0),
\label{eq:Kelstatic}
\end{equation}
is calculated using only the static ($\omega=0$) part of the screened electron-electron interaction $W_{n{\bf{k}},n'{\bf{k}'}}(\omega=0)$, while the dynamical part can be written as
\begin{equation}
\begin{split}
\mathcal{K}^\textrm{el dyn}(\varepsilon,\varepsilon')&=\frac{1}{\tanh(\beta/2 \varepsilon)}\frac{1}{\tanh(\beta/2  \varepsilon')}\frac{1}{\beta}\sum_\nu\left[\frac{1}{N(\varepsilon)N(\varepsilon')}\sum_{n{\bf{k}},n'{\bf{k}'}}\delta(\varepsilon-\varepsilon_{n{\bf k}})\delta(\varepsilon'-\varepsilon_{n {\bf k'}}) W^\text{dyn}_{n{\bf{k}},n'{\bf{k}'}}(i\nu) \right] \\ 
&\times\left( \frac{2( \varepsilon- \varepsilon')}{( \varepsilon- \varepsilon')^2+\nu^2}(n_F( \varepsilon)-n_F( \varepsilon'))+\frac{2( \varepsilon+ \varepsilon')}{( \varepsilon+ \varepsilon')^2+\nu^2}(n_F(- \varepsilon)-n_F( \varepsilon')) \right).
\end{split}
\label{eq:Keldynamic}
\end{equation}
Here, we defined $W^\text{dyn}_{n{\bf{k}},n'{\bf{k}'}}(i\nu)=W_{n{\bf{k}},n'{\bf{k}'}}(i\nu)-W_{n{\bf{k}},n'{\bf{k}'}}(0)$, and the averaging over energy surfaces in Eqs.~\eqref{eq:Kelstatic} and~\eqref{eq:Keldynamic} is done numerically using the tetrahedron method with the electronic states $n$ limited to the low-energy model window described in the next section.

\section{Computational details for the first principles calculations}

For the DFT calculations in \textsc{Quantum} ESPRESSO \cite{Giannozzi2009,Giannozzi2017} with the GGA functional \cite{Perdew1996}, we use the experimental lattice paremeters for space group $I4/mmm$ (139) as listed in Table~\ref{tab:parameters} and employ the PAW pseudopotentials (PPs) from the PSlibrary \cite{DalCorso2014}. We use energy cutoffs 50 Ry (440 Ry) for the wave-functions (charge density) in systems where $f$ states are not considered in the valence window, and 100 Ry (600 Ry) for the Th- and La-based compounds where the unoccupied $f$ states are close to the Fermi energy.

We calculate the electron-phonon matrix elements $g_{\lambda_{\bf q}}^{n{\bf k},n'{\bf k'}}$ on a $6\times6\times6$ ${\bf q}$-grid and a $12\times12\times12$ ${\bf k}$-grid and use interpolated grids of $24\times24\times24$ ($\bf{q}$) and $60\times60\times60$ ($\bf{k}$) in the integration of the electron-phonon coupling constant $\lambda$ and Eliashberg function $\alpha^2F(\omega)$ for convergence. While even finer grids could result in some changes to our results, these choices are a tradeoff between numerical cost and accuracy and are consistent with previous phonon calculations for the quaternary borocarbides by other groups \cite{Weber2014,Tutunucu2015,Uzunok2015} (which also showed good agreement with experiment for YNi$_2$B$_2$C \cite{Weber2014}).

To obtain the electron-electron interactions, we first perform a DFT calculation using the FLEUR full-potential all-electron code \cite{fleurcode} and define a low-energy model using maximally localized Wannier functions \cite{Marzari1997,Mostofi2008}, starting from initial $R$--$d$, $M$--$d$, B--$p$ and C--$p$ projections (in the systems where we consider $f$ states we also include an initial $R$--$f$ projection). Due to the use of a large model with 24 (31 with $f$ states) Wannier functions we can accurately reproduce the low-energy region from at least $\sim-5$~eV to $\sim5$~eV with resulting Wannier functions that are well localized and atom-centered (or close to atom-centered in some cases). 
We then calculate the corresponding frequency-dependent screened interaction $W_{\bf k}(\omega)$ in this basis within RPA \cite{Pines1963} on a $6\times6\times6$~${\bf k}$-grid  using the SPEX code \cite{Friedrich2010}. 

We also compute the effective bare interaction of the $M$--$d$ states using constrained RPA (cRPA) \cite{Aryasetiawan2004}. This is done by separating the screening channels into a ``$d$" and ``rest" space and calculating
\begin{equation}\label{eq:UcRPA_Pr}
U_{\bf q}(\omega)=[1-v_{\bf q}\Pi_r(\omega,{\bf q})]^{-1} v_{\bf q},
\end{equation}
where $v$ is the bare Coulomb interaction and $\Pi_r$ is the ``rest" polarization, which includes screening channels between states outside of the $d$ manifold and between the $d$ and outside states. Note that screening $U$ with the screening channels from the $d$ states, $\Pi_d$, recovers the screened interaction $W$:
\begin{equation}\label{eq:WRPA}
[1-U\Pi_d]^{-1} U=[1-U\Pi_d]^{-1}[1-v\Pi_r]^{-1} v=[1-v\Pi]^{-1}v=W,
\end{equation}
with $\Pi$ the RPA polarization function.

\section{DFT results}

\begin{figure}
\includegraphics[width=0.97\textwidth]{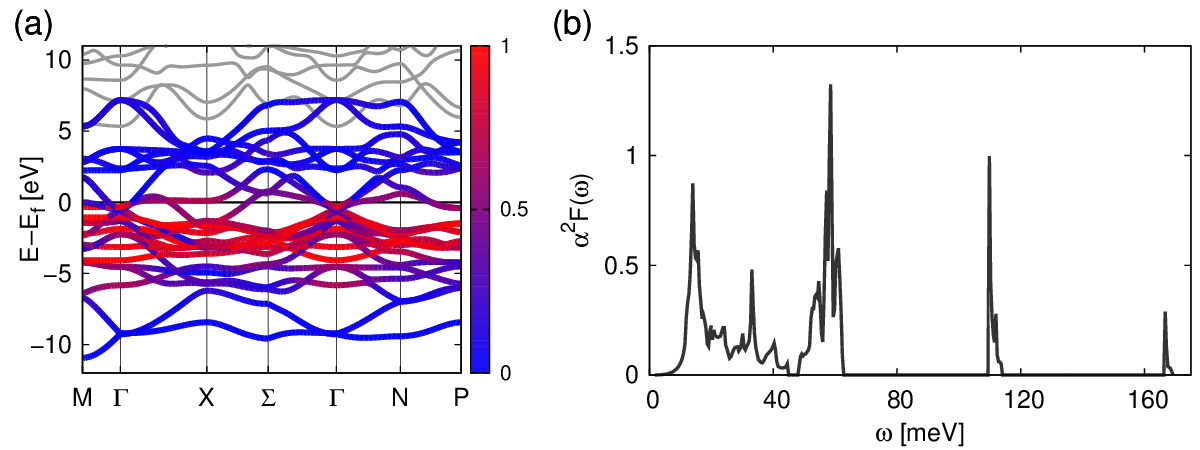}
\vspace{-0.6cm}\caption{ScNi$_2$B$_2$C: (a) DFT and Wannier band structure and (b) Eliashberg function $\alpha^2F(\omega)$.
\label{ScNi}}
\end{figure}

\begin{figure}
\includegraphics[width=0.97\textwidth]{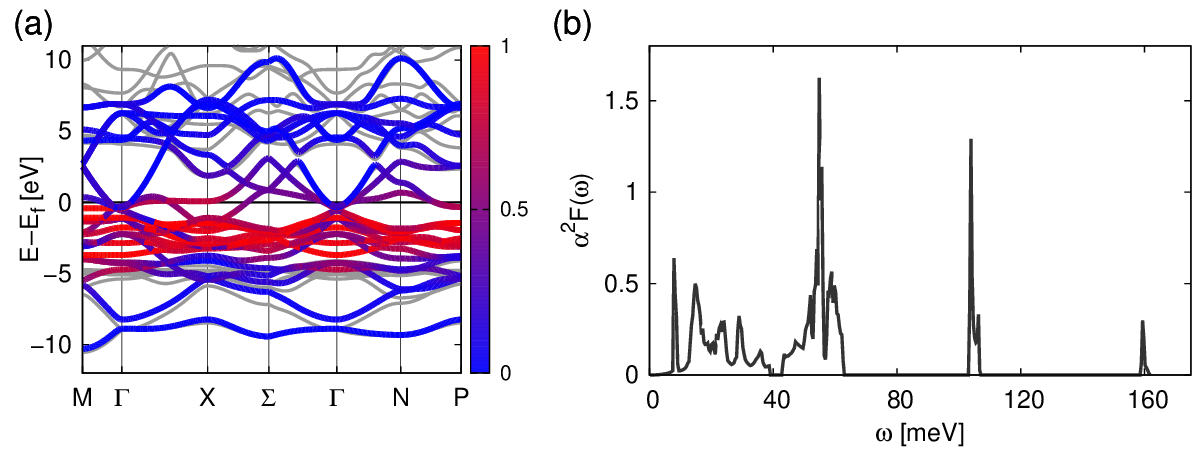}
\vspace{-0.6cm}\caption{LuNi$_2$B$_2$C: (a) DFT and Wannier band structure and (b) Eliashberg function $\alpha^2F(\omega)$.
\label{LuNi}}
\end{figure}

\begin{figure}
\includegraphics[width=0.97\textwidth]{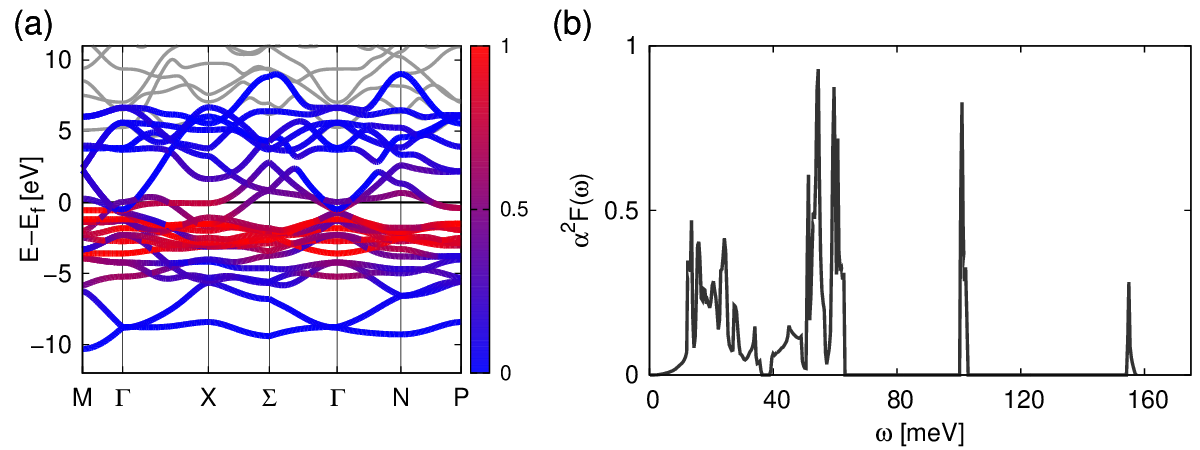}
\vspace{-0.6cm}\caption{YNi$_2$B$_2$C: (a) DFT and Wannier band structure and (b) Eliashberg function $\alpha^2F(\omega)$.
\label{YNi}}
\end{figure}

\begin{figure}
\includegraphics[width=0.97\textwidth]{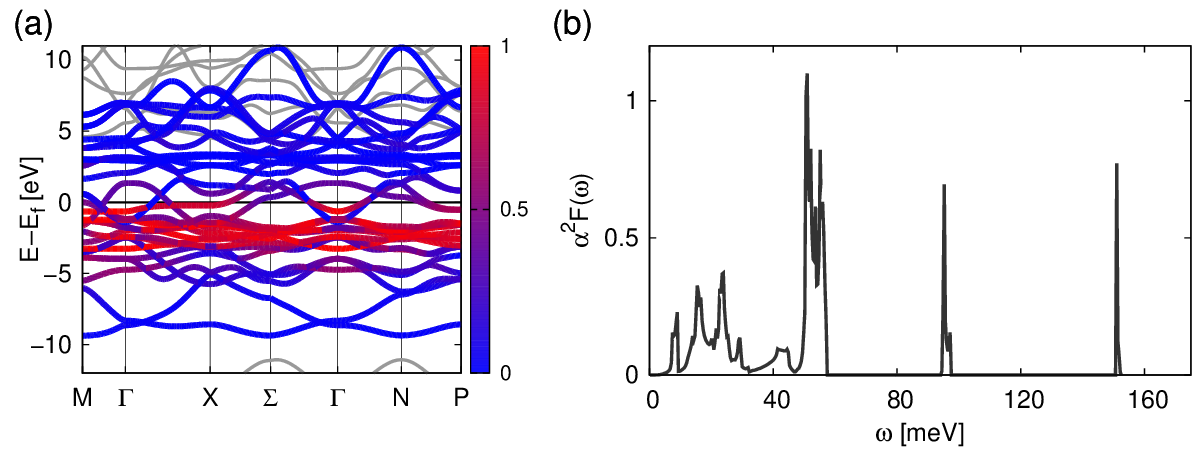}
\vspace{-0.6cm}\caption{ThNi$_2$B$_2$C: (a) DFT and Wannier band structure and (b) Eliashberg function $\alpha^2F(\omega)$.
\label{ThNi}}
\end{figure}

\begin{figure}
\includegraphics[width=0.97\textwidth]{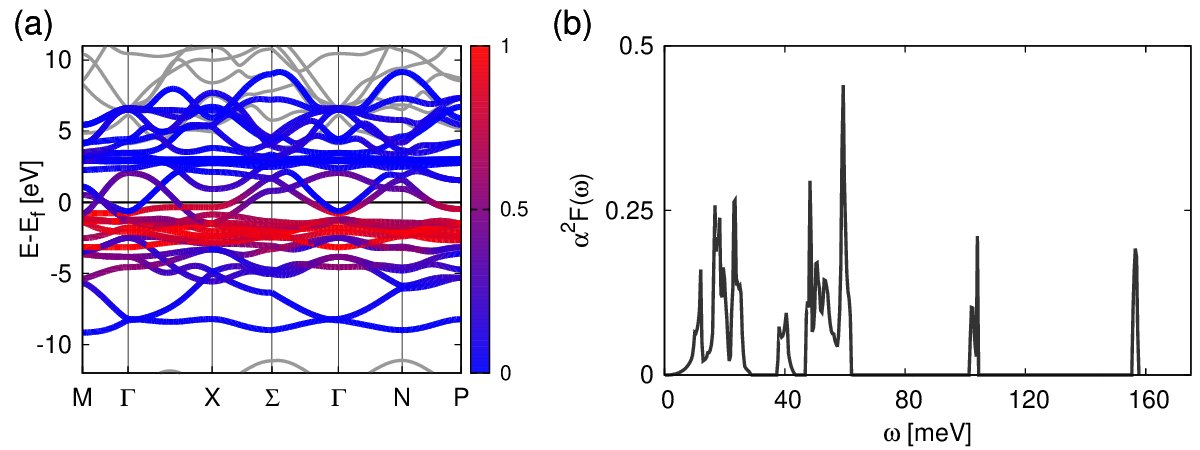}
\vspace{-0.6cm}\caption{LaNi$_2$B$_2$C: (a) DFT and Wannier band structure and (b) Eliashberg function $\alpha^2F(\omega)$.
\label{LaNi}}
\end{figure}

\begin{figure}
\includegraphics[width=0.97\textwidth]{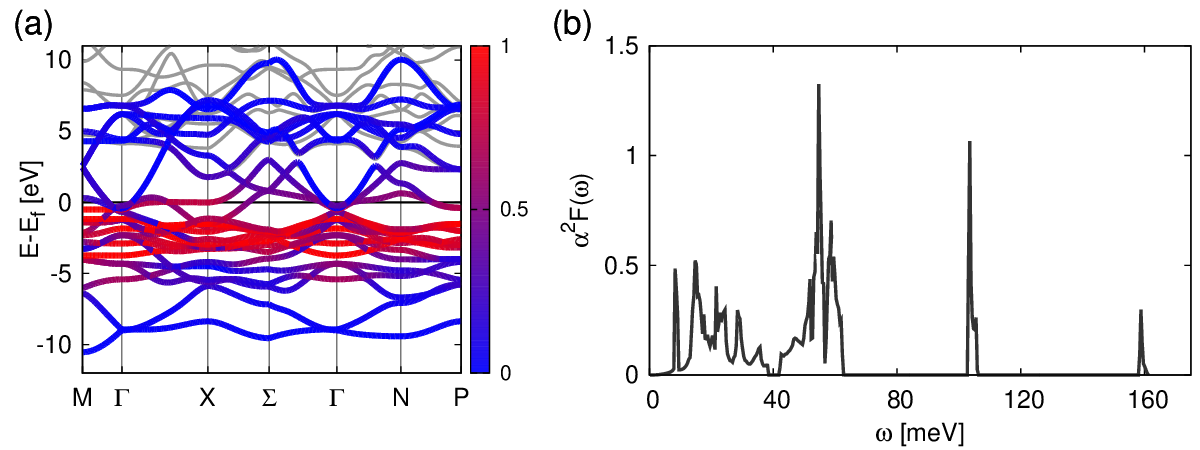}
\vspace{-0.6cm}\caption{YbNi$_2$B$_2$C: (a) DFT and Wannier band structure and (b) Eliashberg function $\alpha^2F(\omega)$.
\label{YbNi}}
\end{figure}

\begin{figure}
\includegraphics[width=0.97\textwidth]{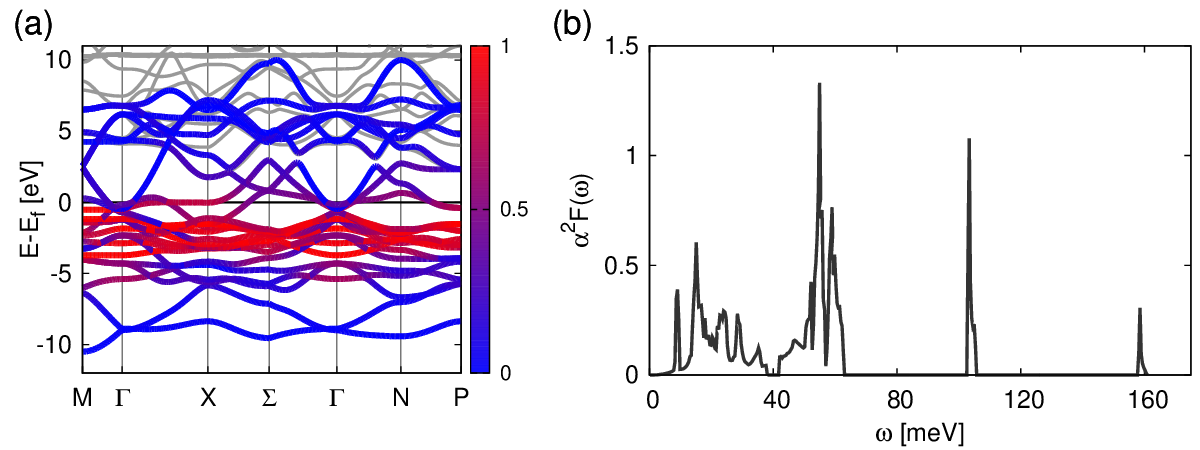}
\vspace{-0.6cm}\caption{TmNi$_2$B$_2$C: (a) DFT and Wannier band structure and (b) Eliashberg function $\alpha^2F(\omega)$.
\label{TmNi}}
\end{figure}

\begin{figure}
\includegraphics[width=0.97\textwidth]{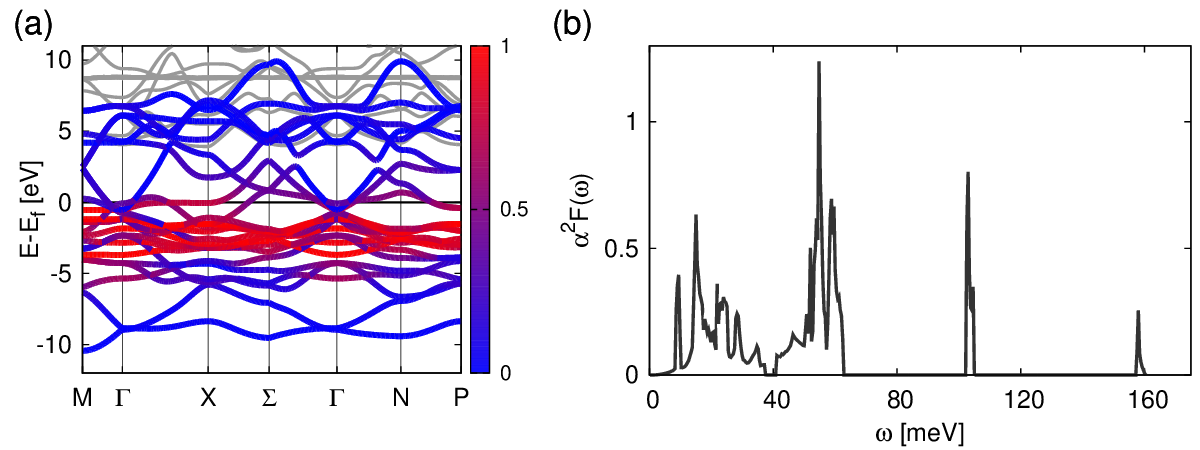}
\vspace{-0.6cm}\caption{ErNi$_2$B$_2$C: (a) DFT and Wannier band structure and (b) Eliashberg function $\alpha^2F(\omega)$.
\label{ErNi}}
\end{figure}

\begin{figure}
\includegraphics[width=0.97\textwidth]{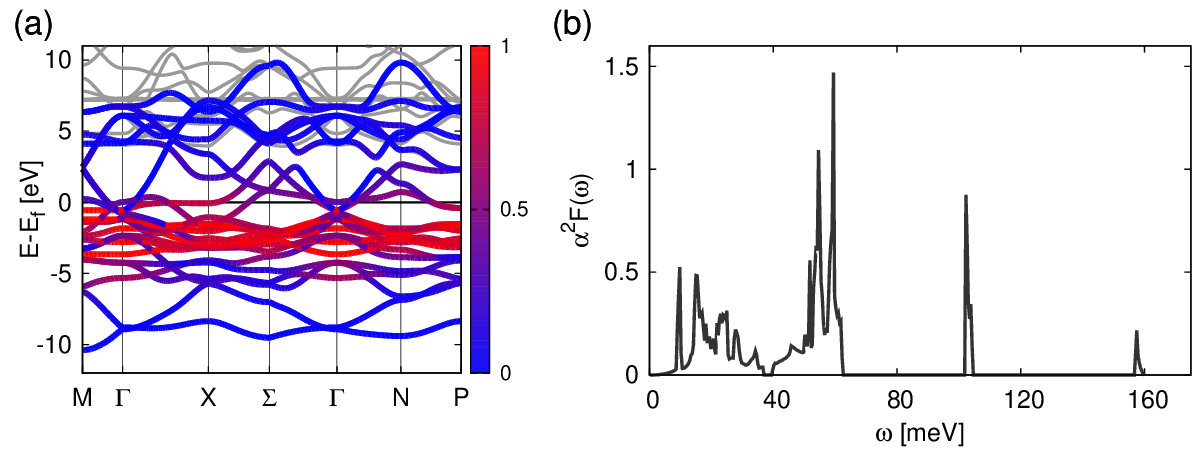}
\vspace{-0.6cm}\caption{HoNi$_2$B$_2$C: (a) DFT and Wannier band structure and (b) Eliashberg function $\alpha^2F(\omega)$.
\label{HoNi}}
\end{figure}

\begin{figure}
\includegraphics[width=0.97\textwidth]{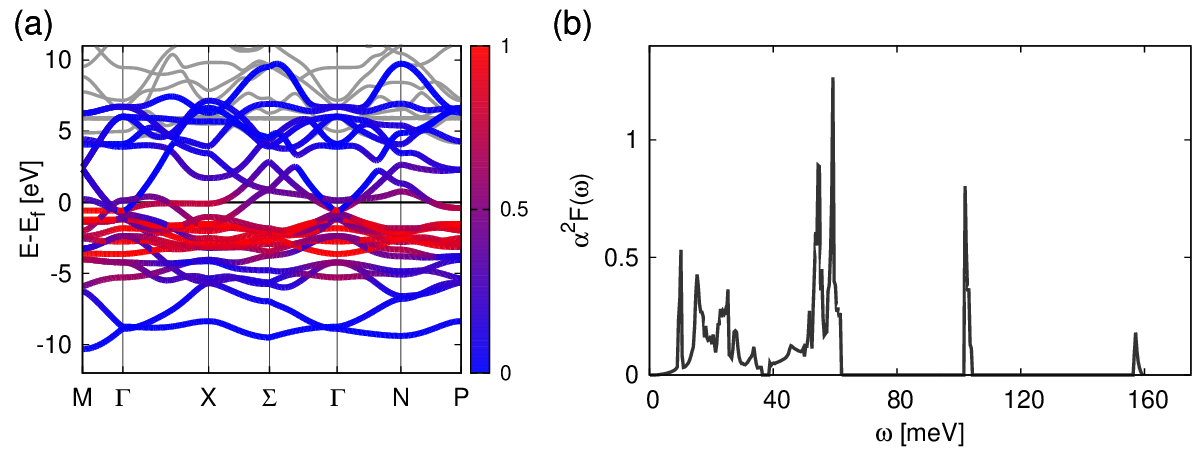}
\vspace{-0.6cm}\caption{DyNi$_2$B$_2$C: (a) DFT and Wannier band structure and (b) Eliashberg function $\alpha^2F(\omega)$.
\label{DyNi}}
\end{figure}

\begin{figure}
\includegraphics[width=0.97\textwidth]{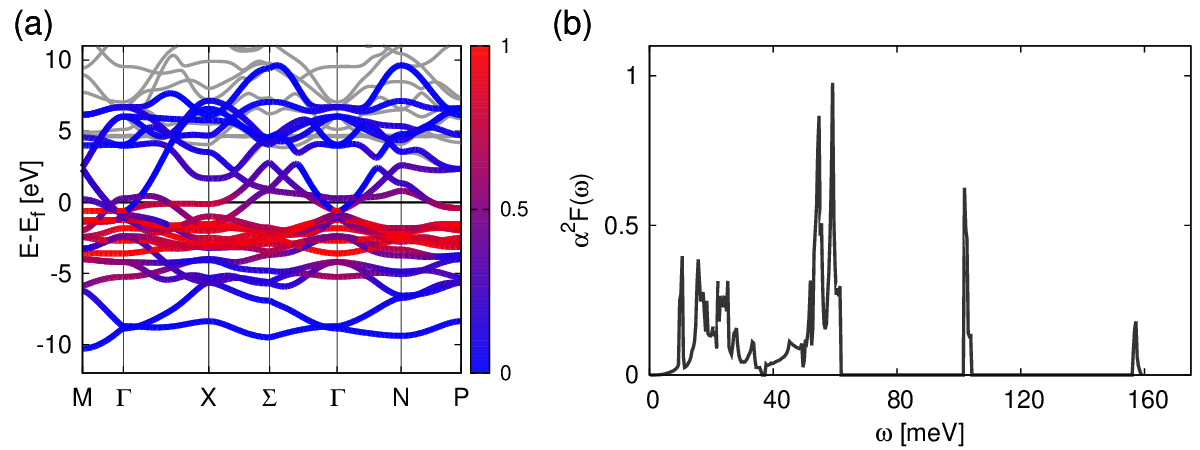}
\vspace{-0.6cm}\caption{TbNi$_2$B$_2$C: (a) DFT and Wannier band structure and (b) Eliashberg function $\alpha^2F(\omega)$.
\label{TbNi}}
\end{figure}

\begin{figure}
\includegraphics[width=0.97\textwidth]{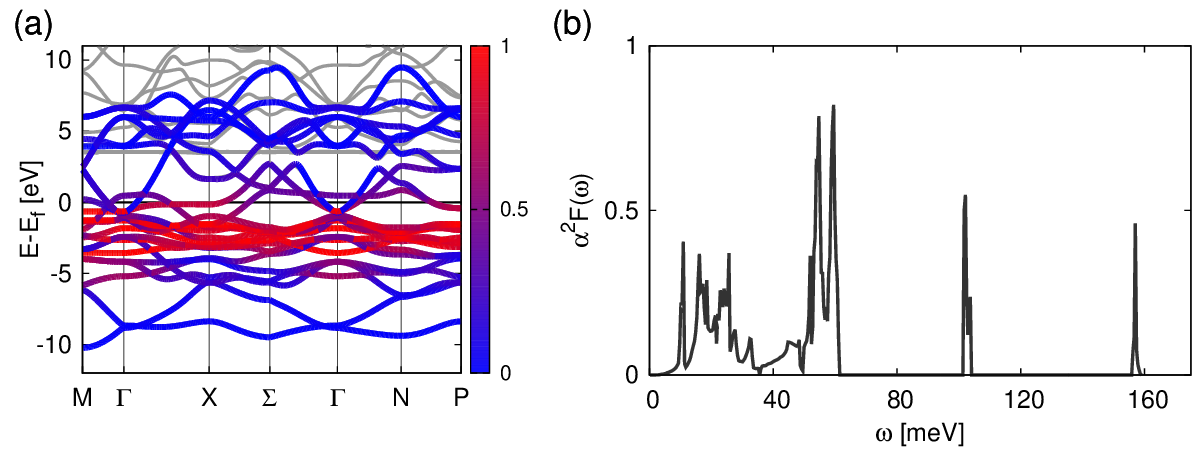}
\vspace{-0.6cm}\caption{GdNi$_2$B$_2$C: (a) DFT and Wannier band structure and (b) Eliashberg function $\alpha^2F(\omega)$.
\label{GdNi}}
\end{figure}

\begin{figure}
\includegraphics[width=0.97\textwidth]{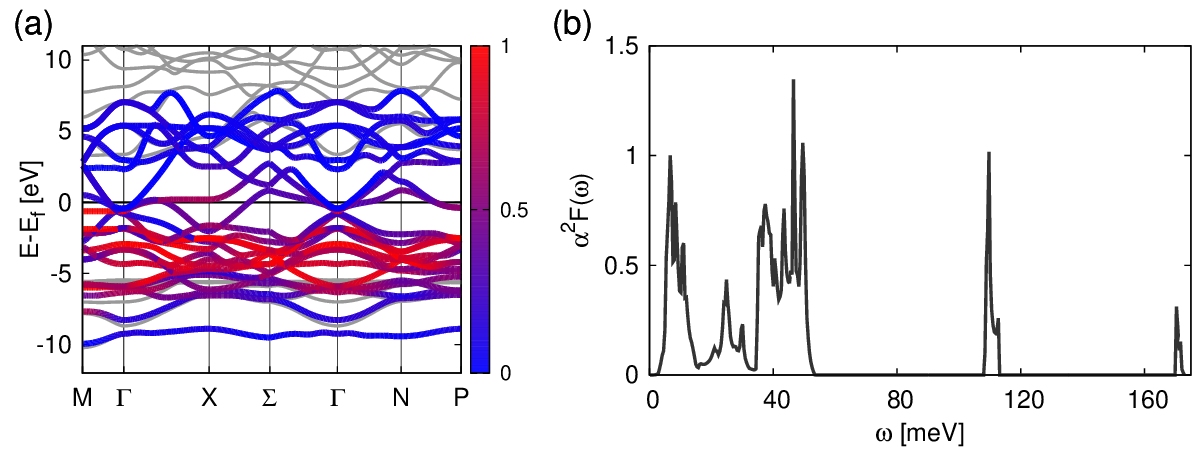}
\vspace{-0.6cm}\caption{LuPd$_2$B$_2$C: (a) DFT and Wannier band structure and (b) Eliashberg function $\alpha^2F(\omega)$.
\label{LuPd}}
\end{figure}

\begin{figure}
\includegraphics[width=0.97\textwidth]{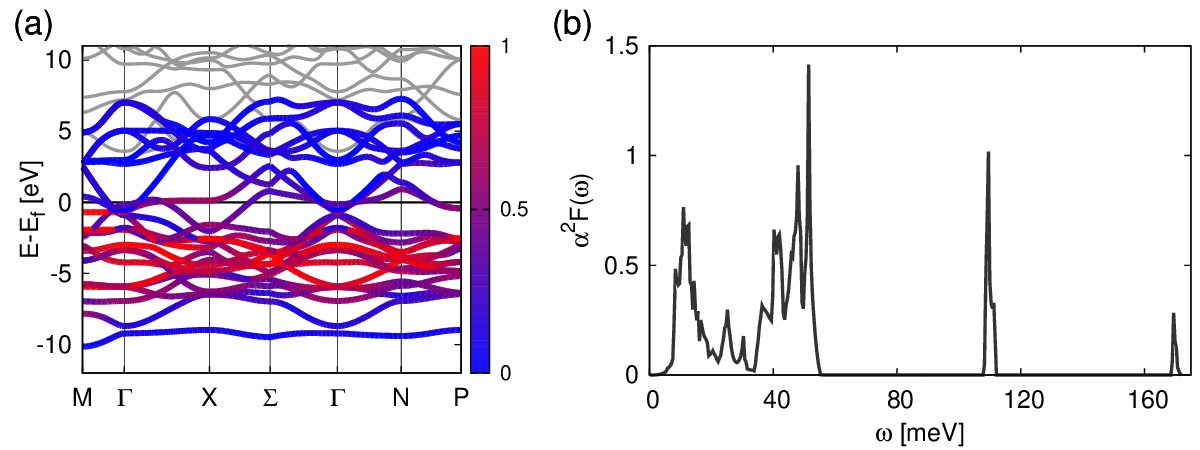}
\vspace{-0.6cm}\caption{YPd$_2$B$_2$C: (a) DFT and Wannier band structure and (b) Eliashberg function $\alpha^2F(\omega)$.
\label{YPd}}
\end{figure}

\begin{figure}
\includegraphics[width=0.97\textwidth]{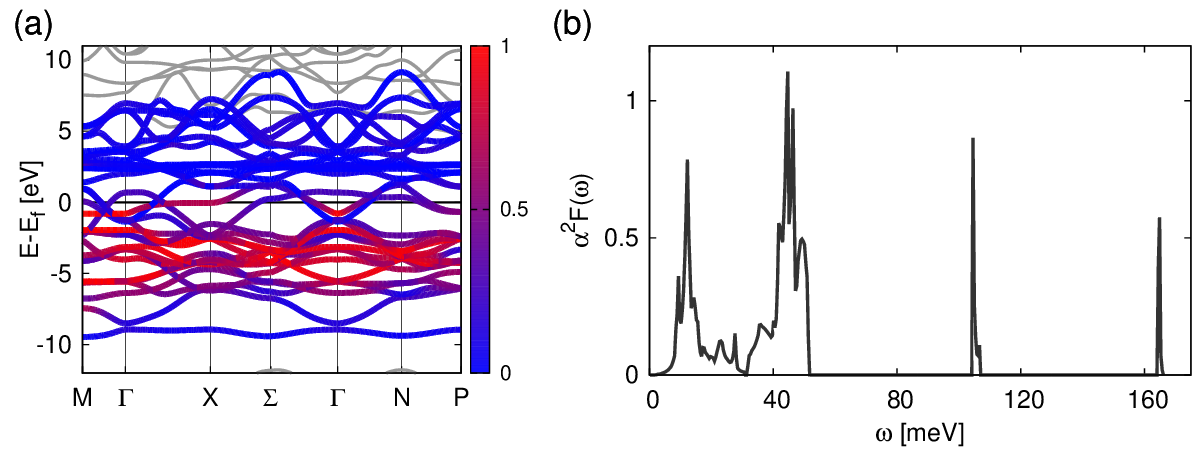}
\vspace{-0.6cm}\caption{ThPd$_2$B$_2$C: (a) DFT and Wannier band structure and (b) Eliashberg function $\alpha^2F(\omega)$.
\label{ThPd}}
\end{figure}

\begin{figure}
\includegraphics[width=0.97\textwidth]{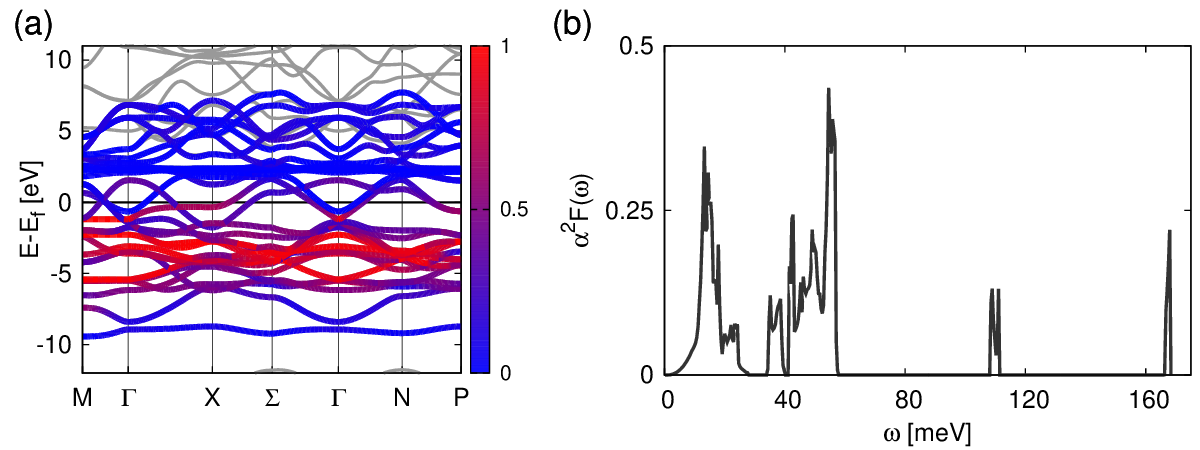}
\vspace{-0.6cm}\caption{LaPd$_2$B$_2$C: (a) DFT and Wannier band structure and (b) Eliashberg function $\alpha^2F(\omega)$.
\label{LaPd}}
\end{figure}

\begin{figure}
\includegraphics[width=0.97\textwidth]{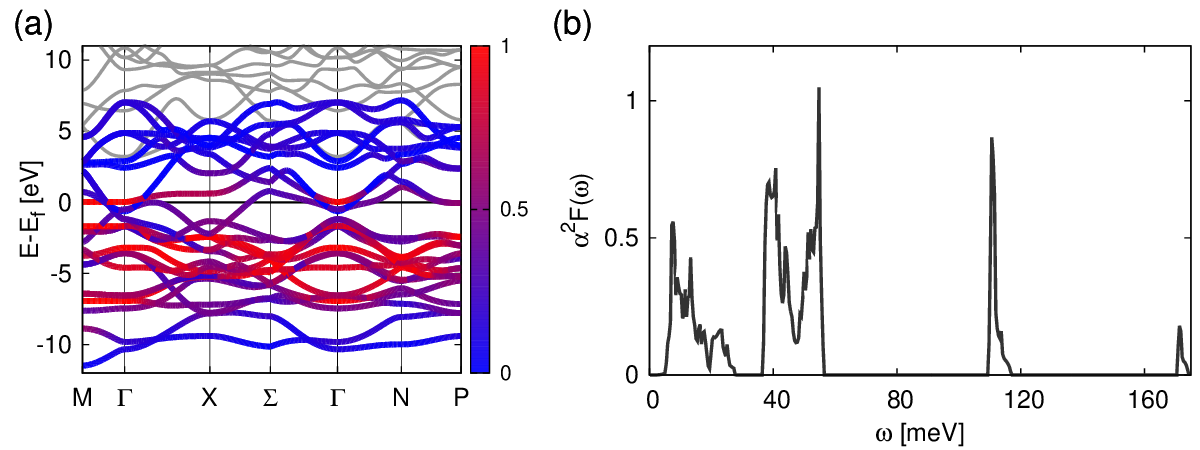}
\vspace{-0.6cm}\caption{YPt$_2$B$_2$C: (a) DFT and Wannier band structure and (b) Eliashberg function $\alpha^2F(\omega)$.
\label{YPt}}
\end{figure}

\begin{figure}
\includegraphics[width=0.97\textwidth]{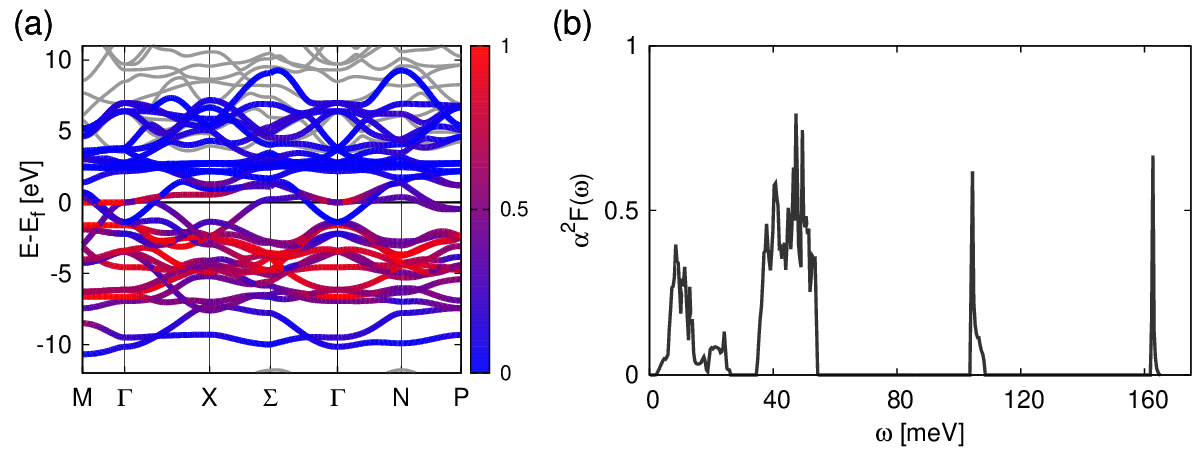}
\vspace{-0.6cm}\caption{ThPt$_2$B$_2$C: (a) DFT and Wannier band structure and (b) Eliashberg function $\alpha^2F(\omega)$.
\label{ThPt}}
\end{figure}

\begin{figure}
\includegraphics[width=0.97\textwidth]{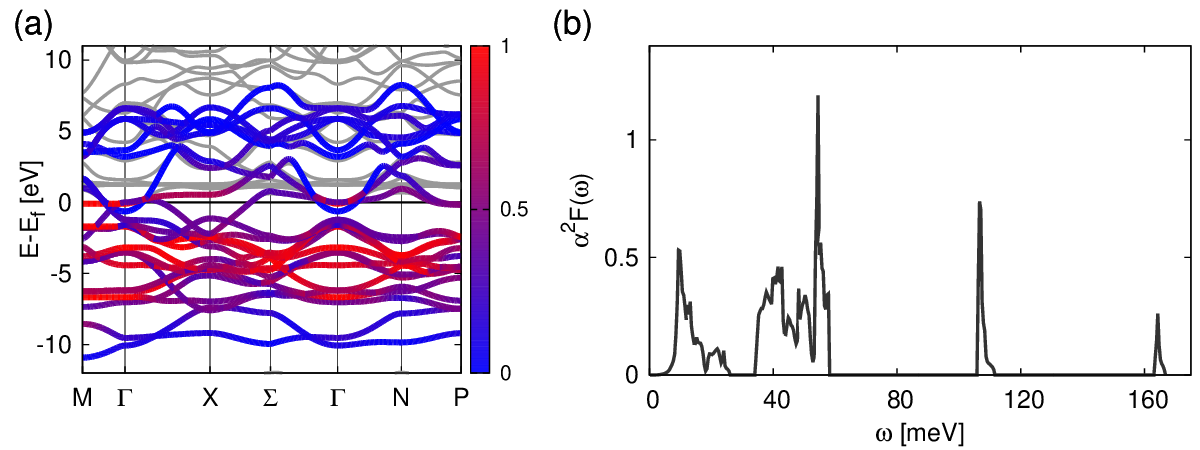}
\vspace{-0.6cm}\caption{PrPt$_2$B$_2$C: (a) DFT and Wannier band structure and (b) Eliashberg function $\alpha^2F(\omega)$.
\label{PrPt}}
\end{figure}

\begin{figure}
\includegraphics[width=0.97\textwidth]{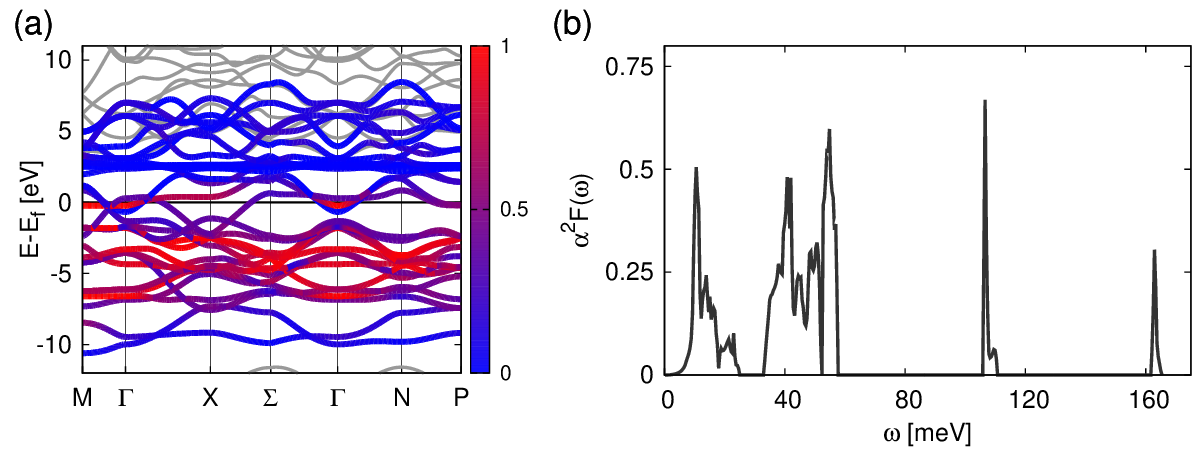}
\vspace{-0.6cm}\caption{LaPt$_2$B$_2$C: (a) DFT and Wannier band structure and (b) Eliashberg function $\alpha^2F(\omega)$.
\label{LaPt}}
\end{figure}

In this section we present the DFT band structures and $\alpha^2F(\omega)$ used in the SCDFT calculations. Figures~\ref{ScNi}-\ref{LaNi} show the non-magnetic Ni-based systems, Figs.~\ref{YbNi}-\ref{GdNi} the magnetic Ni-based systems (with partially filled $4f$ states placed in the core), and Figs.~\ref{LuPd}-\ref{LaPd} the Pd-based and Figs.~\ref{YPt}-\ref{LaPt} the Pt-based compounds. 
The colorbars for the Wannier (disentangled) band structures indicate the $M$--$d$ orbital contributions to the bands.

\subsection{Dynamical instability in YRu$_2$B$_2$C}

\begin{figure}[t]
\includegraphics[width=0.99\textwidth]{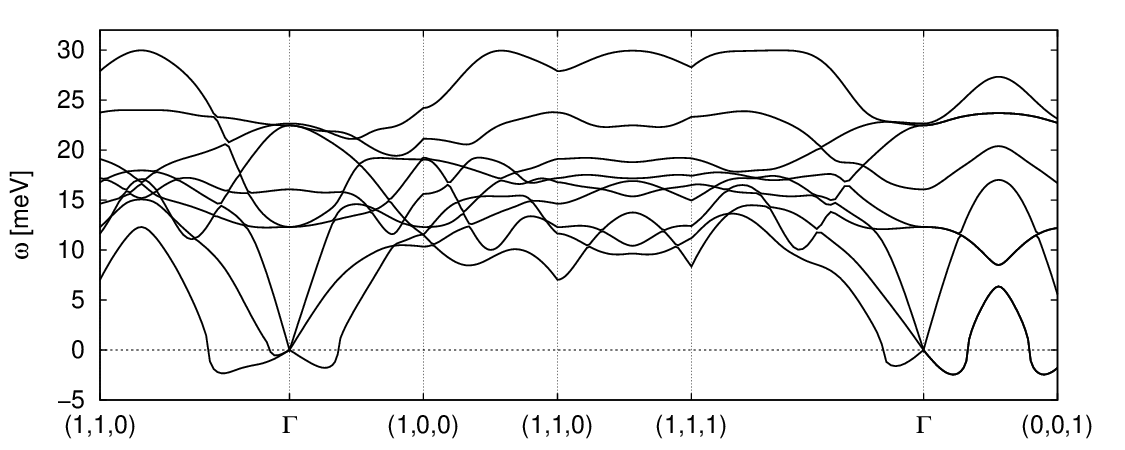}
\caption{Low-frequency part of the phonon dispersion of YRu$_2$B$_2$C along the indicated high-symmetry path with the wave vectors given in reciprocal lattice units $(\frac{2\pi}{a},\frac{2\pi}{a},\frac{2\pi}{c})$. Negative $\omega$ are used to represent imaginary phonon frequencies (soft modes).
\label{YRu_phonon}}
\end{figure}

\begin{figure}[t]
\includegraphics[width=0.99\textwidth]{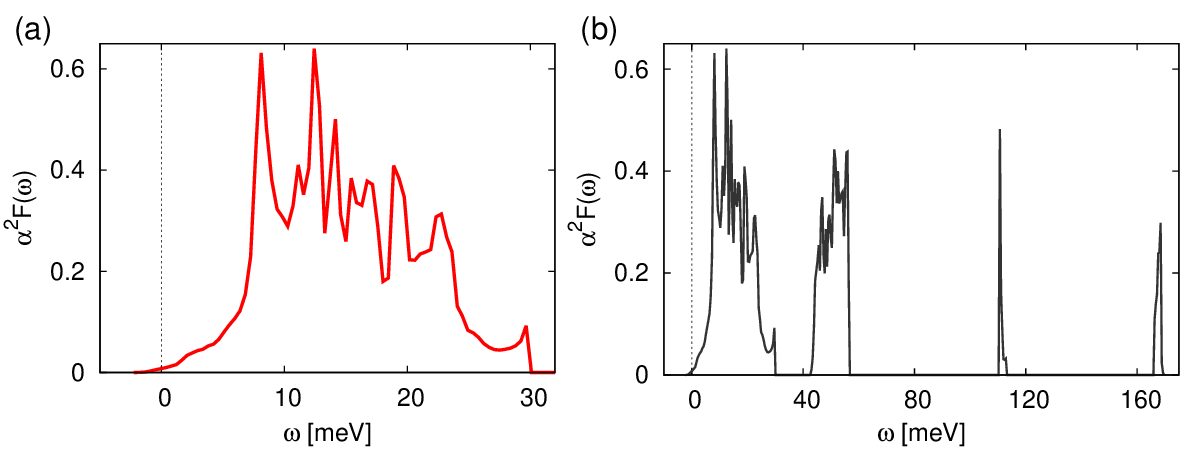}
\caption{YRu$_2$B$_2$C: (a) Eliashberg function $\alpha^2F(\omega)$ in the low-frequency region and (b) in the full window. Negative $\omega$ are used to represent imaginary phonon frequencies (soft modes).
\label{YRu_a2f}}
\end{figure}

As discussed in the main text, the experimentally multiphased YRu$_2$B$_2$C compound is found to be unstable in our phonon calculations. This is indicated by the soft modes in Figs.~\ref{YRu_phonon}  and~\ref{YRu_a2f} (the imaginary phonon frequencies are shown as negative values). The phonon dispersion indicates that the largest instabilities are in the plane along the [110] and [100] directions (note that $a=b$ for space group $I4/mmm$), and along the $c$ direction [001].
As can be seen from $\alpha^2F(\omega)$ in Fig.~\ref{YRu_a2f}, the contribution to the phonon density of states and electron-phonon couplings of these modes is, however, relatively small. To perform the SCDFT calculations reported in the main text we therefore remove these contributions to $\alpha^2F(\omega)$, which enters the calculations of $\mathcal{Z}^\textrm{ph}$ and $\mathcal{K}^\textrm{ph}$.

\newpage

\section{SCDFT results with a static-only electronic kernel}

As discussed in the main text, the $T_c$ calculated using only the static part of the electron-electron interaction ($\mathcal{K}^\textrm{el static}$) are in all cases significantly underestimated when compared to experiment. These values are given in Table~\ref{tab:parameters} ($T_c^\textrm{stat}$) together with the results of the full calculations ($T_c^\textrm{SCDFT}$) and the experimental values ($T_c^\textrm{EXP}$), used for the figures in the main text.

\begin{table}
\caption{Calculated electron-phonon coupling constant $\lambda$ and $\omega_\textrm{ln}$, effective bare local interaction $U_\textrm{cRPA}(\omega=0)$ and screened local interaction $W_\textrm{RPA}(\omega=0)$ in eV for the M--$d$ states calculated within (c)RPA, and calculated critical temperatures using a static electronic kernel ($T_c^\textrm{stat}$)  and the full $\mathcal{K}^\textrm{el}$ ($T_c^\textrm{SCDFT}$).
For the experimental lattice constants ($a$, $c$) and measured $T_c^\textrm{EXP}$ see Ref.~\cite{Niewa2011} and references therein, and Ref.~\cite{Yang1995} for the lattice constants of YPt$_2$B$_2$C. The lattice parameters of LuPd$_2$B$_2$C have been calculated via a full structure relaxation (fixing only the space group).
\label{tab:parameters}
}
\setlength{\tabcolsep}{5.2pt} 
\renewcommand{\arraystretch}{1.2} 
\begin{tabular}{|c| c c | c c | c c c| c c |}
\hline

\hline

& $\lambda$ & $\omega_\textrm{ln}$ [K] & $U$ [eV] & $W$ [eV] & $T_c^\textrm{stat}$ [K] & $T_c^\textrm{SCDFT}$ [K] & $T_c^\textrm{EXP}$ [K] & $a$ [\AA] & $c$ [\AA]  \\

\hline

\hline   
\hline
LaNi$_2$B$_2$C & 0.32 &  270.5  &  2.9 & 2.2 & 0 & $\lesssim 0.4$  & 0 & 3.79  & 9.82   \\

ThNi$_2$B$_2$C & 0.61 &  266.0  &  2.7 & 2.1  & 1.0 & 6.5 & $6-8$  & 3.68 & 10.22    \\

YNi$_2$B$_2$C & 0.74 &  298.5  &  3.0 & 2.1  & 1.9 & 12.5 &  $15.2-15.6$ &  3.53 & 10.54   \\

LuNi$_2$B$_2$C & 0.85 &  266.9  & 3.1 & 2.2  & 4.5 & 14.4  &  $16-16.7$ & 3.46 & 10.63   \\

ScNi$_2$B$_2$C & 0.92 &  270.6  &  3.0 & 2.2  & 4.3 & 13.9  & $15-16$ & 3.35 & 10.68   \\

\hline
\hline

GdNi$_2$B$_2$C & 0.58 &  294.9  & 3.0 & 2.1  & 0.3 & 7.6 &  0  & 3.58 & 10.36  \\

TbNi$_2$B$_2$C & 0.61 & 294.9  &  3.0 & 2.1  & 0.8 & 8.3 &  0  & 3.55 & 10.44 \\

DyNi$_2$B$_2$C & 0.68 &  283.4  &  3.0 & 2.2  & 2.4 & 11.2 & $6-6.4$  & 3.53 & 10.49  \\

HoNi$_2$B$_2$C & 0.76 &  281.4  &  3.0 & 2.1  & 3.4 & 13.9 & $7.5-8.3$  & 3.52 & 10.53  \\

ErNi$_2$B$_2$C & 0.79 &  278.8  &  3.0 & 2.2  & 3.9 & 14.3 & $10.5-11$  & 3.50 & 10.56 \\

TmNi$_2$B$_2$C & 0.81 &  276.1  &  3.0 & 2.2   & 3.9 & 14.4 & $11$ & 3.49 & 10.59  \\

YbNi$_2$B$_2$C & 0.84 &  271.5  &  3.0 & 2.2  & 4.2 & 14.7 &  0 (heavy F)  & 3.48 & 10.61  \\

\hline 
\hline

LaPd$_2$B$_2$C & 0.43 & 247.1 & 2.8 & 2.4  & $\sim 0.1$ & 2.6 &  $1.7-1.9$  & 3.96  & 10.18   \\

ThPd$_2$B$_2$C & 0.90 &  236.0  & 2.7 & 2.3  & 3.4 & 9.9 & $14.4-14.6$  & 3.84 & 10.65  \\

YPd$_2$B$_2$C & 1.35 &  222.2  &  2.8 & 2.4  & 7.7 & 18.9 & $21-23$  & 3.75 & 10.73  \\

LuPd$_2$B$_2$C & 1.93 &  154.7  & 2.8 & 2.5 & 12.4 & 22.8 & --  & 3.73$^*$  & 10.80$^*$  \\

\hline
\hline

LaPt$_2$B$_2$C & 0.72 & 255.4  &  2.6 & 2.2  & 1.5  & 6.1 & $10.5-11$  & 3.87 & 10.71 \\

PrPt$_2$B$_2$C & 0.85 & 241.8 & 2.5 & 2.2 & 2.9 & 8.2 & $6-6.5$  & 3.84 & 10.76 \\

ThPt$_2$B$_2$C & 0.92 &  206.1  &  2.5 & 2.2  & 4.9 & 11.7 &  $6.7-7$  & 3.83 & 10.86 \\

YPt$_2$B$_2$C & 1.13 &  207.4  & 2.6 & 2.2  & 3.6 & 11.8 &  $10-11$  & 3.79 & 10.71 \\

\hline

\end{tabular}
\end{table}

\pagebreak

